\documentclass[11pt]{article}
\newcommand{\be}{\begin{eqnarray}}\newcommand{\beq}{\begin{equation}}
\newcommand{\ee}{\end{eqnarray}}\newcommand{\eeq}{\end{equation}}
\newcommand{\De}{\Delta}
\newcommand{\la}{\lambda}
\input epsf
\input amssym.tex
\usepackage{graphicx} 
\title{Thermodynamics of heterogeneous crystal nucleation in 
contact and immersion modes} 
\author{Y. S. Djikaev\thanks{Corresponding author. E-mail: idjikaev@eng.buffalo.edu}
\hspace{0.1cm} and \hspace{0.1cm}E. Ruckenstein$^{}$\thanks{Phone: 
(716) 645-2911, ext.2214;  Fax: (716) 645-3822; E-mail: feaeliru@acsu.buffalo.edu  }\hspace{0.2cm}  \\ 
\\ Department of Chemical and Biological  Engineering, SUNY at Buffalo, \\ 
Buffalo, New York  14260
\\.\\ 
}
\renewcommand{\baselinestretch}{2}
\begin{document}
\renewcommand{\baselinestretch}{1}
\maketitle
\renewcommand{\baselinestretch}{1}
\begin{abstract}
\renewcommand{\baselinestretch}{1} 
One of most intriguing problems of heterogeneous crystal nucleation in droplets is  its strong
enhancement in the contact mode (when the foreign particle is presumably in some kind of {\em
contact} with the droplet surface) compared to the immersion mode (particle {\em immersed} in the
droplet).  Many heterogeneous centers have different  nucleation thresholds when they act in
contact or immersion modes,  indicating that the  mechanisms may be actually different for the
different modes.  Underlying physical reasons for this enhancement have remained largely  unclear. 
In this paper we present a model for the thermodynamic enhancement of heterogeneous crystal
nucleation in the  contact mode compared to the immersion one. To determine if and how the surface
of a liquid droplet  can  thermodynamically  stimulate its heterogeneous crystallization, we
examine crystal nucleation in the immersion and contact modes by  deriving and comparing  with each
other the reversible works of formation of crystal nuclei in these cases.   As a numerical
illustration, the proposed model is applied to the heterogeneous  nucleation of Ih crystals on
generic macroscopic foreign particles in water  droplets at $T=253$ K. Our results show that the
droplet surface does thermodynamically favor the contact mode over the immersion one. Surprisingly,
our numerical evaluations suggest that the line tension contribution to this enhancement from the
contact of three  water phases (vapor-liquid-crystal) may be of the same order of magnitude as or
even larger than the surface tension contribution. 

\end{abstract} 
\renewcommand{\baselinestretch}{2} 
\newpage 
\section{Introduction}
\renewcommand{\baselinestretch}{2} 
\par The size, composition, and phases of aerosol and cloud particles affect the 
radiative 
and chemical properties$^{1,2}$ of clouds and hence have a great impact on 
Earth's climate as a whole. On the other hand, the composition, size, and phases of
atmospheric particles are determined by the rate at  and mode in which these 
particles form and evolve.$^{2-4}$

Water constitutes an overwhelmingly dominant chemical species
that participates in atmospheric  processes. Consequently, great importance is attributed to studying aqueous aerosols and cloud droplets as well as their phase transformations. 
In a number of important cases 
atmospheric particles appear to freeze homogeneously.$^{4-6}$ For example,
the conversion of supercooled water droplets into ice at temperatures below about
-30$^{o}$C is known to occur homogeneously, mainly because the
concentrations of
the observed ice particles in the clouds often exceed the number
densities of preexisting particles capable of nucleating ice.$^{4,5}$
Also, it has been suggested that aqueous nitric
acid-containing cloud droplets in the polar stratosphere freeze into nitric
acid hydrates via homogeneous nucleation.$^{6}$
Understanding how nitric acid clouds form and grow in the stratosphere is
a topic of current interest because such clouds participate in the heterogeneous
chemistry that leads to springtime ozone depletion over the polar
regions.$^{2}$

However, most phase transformations in aqueous 
cloud droplets occur as a result of heterogeneous nucleation
on preexisting macroscopic particles, macromolecules, or even ions.$^{3}$ 
Heterogeneous nucleation of ice on a microscopic foreign particle can be considered as  
the adsorption of water molecules on a substrate which serves as a template. If the water molecules   
adsorb in a configuration close enough to the crystalline structure of ice, then 
the energy barrier between phases is substantially 
reduced. Recent work on the heterogeneous nucleation of
ice in the atmosphere is motivated by the evidence, primarily
from modeling studies, that heterogeneous freezing may significantly impact the radiative
properties, both in the visible and infrared, of cirrus 
clouds. 
The leading candidates for heterogeneous nucleating centers 
are the mineral dusts (fly ash and metallic particles) and emissions from
aircraft, primarily soot.$^{7-9}$ 
Interest in cirrus clouds motivated several laboratory studies as well.$^{10,11}$ 
They showed that the presence of various foreign  
inclusions shifts the apparent freezing temperature of droplets 
upward by as much as 10$^\circ$C. 

Most investigators targeted particulates  as the primary heterogeneous ice nucleating centers  in
the atmosphere.  Recently, however, increasing attention is payed to the role of films of
high-molecular-weight organic compounds located on droplets. Such compounds are emitted into the atmosphere,
especially in regions that are influenced by biomass burning.$^{12}$ It was reported,
for example, that the films of  long-chain alcohols and some other organic species can  catalyze
ice nucleation in droplets at a  supercooling of only 1$^\circ$C.$^{13,14}$ 

So far, the physical mechanism underlying  heterogeneous crystal nucleation in droplets remains
rather obscure.$^{2}$  As an additional mystery,  many heterogeneous centers have different
nucleation thresholds when they act in different modes - contact or immersion,  indicating that
the  mechanisms may be actually different for the different modes. In the contact mode, the
ice-nucleating particle contacts water droplet, i.e., touches or intersects its surface.  whereas
in the immersion  mode the particle is immersed in the water droplet (Figure 1).$^{2,15}$
The same particle tends to trigger the freezing of  an supercooled water droplet at a higher
temperature in the contact  mode than in the immersion one.$^{2,15,16}$ 

The cause of this enhancement is unknown, but it provides a hint that the water 
surface could be of special interest in ice nucleation.  Several investigators have put forward
conjectures on the mechanism of contact nucleation, all of which depend on the contact of a particle 
impinging upon the droplet surface from air.$^{2}$ One of the hypothesis is based on the partial solubility of small solid particles whereby active sites at the surface of a particle are subject to erosion after it becomes immersed in water.$^{17,18}$ Another hypothesis suggests$^{19}$ that only those particles  enhance nucleation in the contact mode which exhibit a strong affinity for water. During the initial contact with the droplet ({\em before} the equilibrium adsorption is achieved) such particles might strongly lower the free energy barrier to ice nucleation at its surface. Another interesting explanation$^{20}$ suggests that the contact mode enhancement of crystal nucleation is due to the mechanically forced rapid spreading of water along the hydrophobic solid surface which forces its local wetting and thereby temporarily creates local high interface-energy zones increasing the  probability of crystal nucleation.
While acceptable for some particular cases, all those explanations have some inconsistencies and limitations,  and so far no rigorous (and general enough)  theoretical model of this phenomenon has been proposed. 

As a related problem,  recently a  thermodynamic theory was developed$^{21,22}$  that  prescribes the condition 
under which the  surface of a droplet can stimulate {\em homogeneous} crystal nucleation therein so
that  the homogeneous formation of a crystal nucleus with one of its facets at the droplet surface (surface-stimulated mode) 
is thermodynamically favored over its formation with all the facets {\em within} the liquid phase (volume-based mode). 
For both  unary and multicomponent droplets  the inequality
coincides with the  condition for the partial wettability of at least one of the facets of a
crystal nucleus by  its own melt.$^{23}$ This effect was experimentally observed for  several
systems,$^{24,25}$ including water-ice$^{26}$ at temperatures at or  below 0$^{o}$C. 

\par Clearly, the mode of crystal nucleation is most likely determined by both thermodynamic and kinetic factors. However,  the partial wettability of a solid by its melt
may help to explain why, in molecular dynamics simulations$^{}$ of 
various kinds of supercooled liquid droplets$^{27,28}$ 
the crystal nuclei appear preferentially close to the surface. 
Since smaller droplets have a higher surface-to-volume ratio, the per-droplet nucleation rates
in small droplets tend to be higher than in the bulk. Hence  it is experimentally easier to
observe the crystallization of aerosols,  having a large collective surface area, than those
having a large volume. 
Recent experiments$^{16}$  on the heterogeneous freezing of water droplets in both
immersion and contact modes  have also provided evidence that the rate of crystal nucleation in
the contact mode is much higher because the droplet surface may stimulate 
heterogeneous  crystal nucleation in a way similar to the enhancement of the homogeneous
process. 

In this paper we extend the approach, previously developed in refs.21,22,  
to heterogeneous crystal nucleation on a solid 
particle (in both immersion and contact modes) and  present a thermodynamic model thereof.  Our
thermodynamic analysis suggests that, indeed, the droplet surface can 
thermodynamically  enhance crystal nucleation in the contact mode compared to the immersion
mode. Whether this occurs  or not for a particular foreign particle is determined, however, by the
interplay between  five surface tensions and four line tensions involved in this process. 

\par The paper is structured as follows. In section 2 we derive and compare with each other 
the expressions for
the free energy of heterogeneous formation of a crystal nucleus on a solid (say, dust)
particle in the immersion and contact
modes. For the sake of simplicity, in this work we consider only unary
systems, i.e., pure water droplets, but the generalization to  multicomponent droplets can be 
carried out as well. 
Only one kind of  foreign nucleating centers is considered, namely, those
completely wettable by water. Numerical 
predictions and possible experimental verification of the model are discussed in Section 3.
The results and conclusions are summarized in section 4.  

\section{Free energy of heterogeneous formation of crystal nuclei in contact and immersion modes}

To determine if and how the surface of a liquid droplet can thermodynamically  stimulate its
heterogeneous crystallization, it is  necessary to consider the formation of 
a crystal cluster in the two modes (Figure 1). 
In the ``immersion" mode, the crystal cluster is formed 
with one of its facets on a foreign particle that is completely immersed in a liquid droplet; all
other crystal facets interface the liquid. In the ``contact" mode, the foreign particle touches
(i.e., is in contact with) the droplet surface and 
the cluster forms with one of the crystal 
facets on the particle (as in the immersion mode), another facet 
at the liquid-vapor interface, and all other facets making the ``crystal-liquid" interface. 
In these two cases the reversible works of formation of a crystal nucleus
(critical cluster) should be derived and compared with each other. 
This can be carried out in the framework of the classical nucleation theory (CNT) for both
unary and multicomponent droplets. 
In this paper we consider the crystallization of unary droplets. 

\par The droplet surface can incur some  deformation if its 
crystallization is initiated at its surface. The thermodynamic analysis of the case where the
crystallization begins at a droplet surface can be considerably more complicated when compared to
the case where it forms at the surface of a bulk liquid. However, one can show$^{21}$  that if
$a^{}_{1}/\pi R^{2}\ll 1$ (where $a^{}_{1}$ is the surface area of the crystal facet
interfacing the vapor and $R$ is the droplet radius), the formation of a crystal at a droplet
surface can be considered as crystallization at the surface of a {\em bulk} liquid. Under
conditions relevant to the freezing of atmospheric droplets, crystal nuclei are usually of sub- or
nanometer size, while the droplets themselves are in submicron to micrometer size range, i.e., the 
above condition is well satisfied. Since the analysis of the  freezing of atmospheric droplets is
our ultimate goal, one can assume the droplet surface to be flat and thus avoid the complexity of
taking into account the droplet deformation upon freezing.  Besides, heterogeneous particles
serving as nucleating centers can be considered as macroscopic particle of linear sizes much
greater than the crystal nuclei hence the part of its surface on which the crystal nucleus forms can
be considered to be flat as well. 

\par Let us consider a single-component bulk liquid. A macroscopic heterogeneous particle
is either completely immersed in the liquid or in contact with the liquid-vapor interface. 
Crystallization will take place in this liquid if it is in a metastable
(supercooled) state.  The reversible work of crystal formation, $W$, can be found as 
the difference between $X_{fin}$, the 
appropriate thermodynamic potential of the system in its final state
(liquid+crystal), and $X_{in}$, the same potential in its initial state
(liquid): $W=X_{fin}-X_{in}$. 
Since the density of the liquid may be different from that of the solid, the
volume of the liquid may change upon crystallization if the process is not constrained to be
conducted at constant volume. In such a case, strictly speaking, one cannot calculate $W$ as the
difference in the Helmholtz free energies since the volume work that the entire system exchanges
with the environment should not be regarded as work involved in the formation of a local nucleus.
As an approximation, the use of the Helmholtz free energy is still acceptable since, in the
thermodynamic limit, the change in the total volume of the system is usually negligible. A better
choice for the thermodynamic potential is the Gibbs free energy if the system is in contact with a
pressure reservoir (since the unconnected volume work exchanged with the environment is
automatically removed from the Gibbs free energy). However,$^{29}$ in the thermodynamic limit, 
the use of either the Gibbs or Helmholtz free energy or grand 
thermodynamic potential is acceptable for the evaluation of $W$. 

\par Neglecting the density difference between liquid and
solid phases and 
assuming the crystallization process to be isothermal, one can say
that the total volume, the temperature, and the number of molecules  in the system,
respectively, will be constant.
Thus the reversible work of formation of a crystal embryo can be
evaluated as the difference between $F_{fin}$, the
Helmholtz free energy of the system in its final state (liquid+crystal+foreign particle),
and $F_{in}$, in its initial state (liquid+foreign particle): 
\beq W=F_{fin}-F_{in}.\eeq

\subsection{Foreign particle completely immersed in the liquid }

\par Consider a bulk liquid in a container whose upper surface is in contact with the vapor phase
of constant pressure and temperature. A macroscopic foreign particle is completely immersed in this
liquid.  Clearly, for this system to be in mechanical and thermodynamic equilibrium, the
particle must be completely wettable by the liquid. Upon sufficient supercooling, a crystal nucleus
may form heterogeneously with one of its facets on the foreign particle. The crystal is considered
to be of arbitrary shape with $\lambda$ facets (Figure 2).  We will assign the subscript
``$\lambda$'' to the facet which is in contact with the foreign particle.(Figure 3) 

Let us introduce the superscripts
$\alpha,\; \beta,\;\gamma$ and  $\delta$ to denote quantities in the
liquid, vapor, crystal nucleus, and foreign particle, respectively. Double
superscripts will denote quantities at the corresponding interfaces, and triple superscripts at the corresponding
three-phase contact lines.
The surface area and 
surface tension of facet $i\;\;\;(i=1,...,\la)$ will be denoted by $A_{i}$ and
$\sigma_{i}$, respectively. (Anisotropic interfacial
free energies are believed to be particularly important in determining the
character of the nucleation process.) Hereafter, we adopt
the definition of the surface tension of a solid, $\sigma^{solid}$, as given in
chapter 17 of ref.19. Namely,
$\sigma^{solid}=f'+\sum_i \Gamma_i\mu'_i$, where
$f',\;\;\Gamma,$ and $\mu'$ are the surface free energy per unit area,
adsorption, and chemical potential of component $i$, all attributed to the dividing
surface between solid and fluid. In the following, we will neglect the adsorption at
the solid-fluid interfaces. Thus, by
definition, the surface tension of the solid will be equal to the surface free energy
per unit area. 

\par Let us denote the number of molecules in the crystal cluster by $\nu$.  Neglecting the density change
upon freezing and assuming the equality of pressures in the vapor and liquid,  $P^{\alpha}=P^{\beta}$, the 
reversible  work of heterogeneous formation of the crystal (with its facet $\la$ on the foreign particle)
is given by the expression 
\beq W^{imm}=\nu[\mu^{\gamma}(P^{\gamma},T)-\mu^{\alpha}(P^{\alpha},T)]-
V^{\gamma}(P^{\gamma}-P^{\alpha})
+\sum_{i=1}^{\lambda-1}\sigma^{\alpha\gamma}_{i}A^{\alpha\gamma}_{i}
+\sigma_{\la}^{\gamma\delta}A_{\la}^{\gamma\delta}
-\sigma^{\alpha\delta}A_{\la}^{\gamma\delta}+\tau^{\alpha\gamma\delta}L^{\alpha\gamma\delta},\eeq
where $\mu,\;P,\;V$, and $T$ are the chemical potential, pressure, volume, and temperature,
respectively, and $\tau$ is the line tension associated with a three-phase contact line$^{30}$ of length $L$. 

The necessary and sufficient conditions for the equilibrium shape (known as the Wulff form)  of
the crystal are represented by a series of equalities referred to as Wulff's relations (see, e.g.,
ref.23), which  can be regarded as a series of equilibrium conditions on the crystal
``edges'' formed by adjacent facets. For example, on the edge between homogeneously formed 
facets $i$ and $i+1$ the equilibrium condition is 
\beq
\frac{\sigma^{\alpha\gamma}_{i}}{h_{i}}=
\frac{\sigma^{\alpha\gamma}_{i+1}}{h_{i+1}}\;\;\;(i=1,\ldots,\la),\eeq
where $h_{i}$ is the distance from facet $i$ to a point $O$ within the crystal 
(see Figure 2) resulting from the Wulff construction.$^{23}$  
\par In the case when one of the facets (facet $\la$)
is the crystal-vapor interface while
all the others lie within the liquid phase (see Figure 3), the
equilibrium conditions on the edges formed by this facet with the
adjacent ones (hereafter marked by a subscript $j$) are given by
\beq \frac{\sigma^{\alpha\gamma}_{j}}{h_{j}}=
\frac{\sigma^{\gamma\delta}_{\la}-\sigma^{\alpha\delta}}{h_{\la}}.\eeq
Note that the height of the $\la$-th pyramid
(constructed with the base on facet $\la$ and with the apex at point $O$
of the Wulff crystal) will differ from that with all of
the facets in the liquid. Thus, the shape of the crystal
will differ from that in which all facets are in contact with the
liquid. For this case, Wulff's relations take the form
\beq \frac{\sigma^{\alpha\gamma}_{1}}{h_{1}}=
\frac{\sigma^{\alpha\gamma}_{2}}{h_{2}}=\ldots
=\frac{\sigma^{\alpha\gamma}_{\lambda}-\sigma^{\alpha\delta}}{h_{\lambda}
}\eeq
(see also refs.21,22). 
In the above consideration, it is assumed that the mechanical effects
within the crystal (e.g., stresses) reduce to an isotropic pressure
$P^{\gamma}$. In this case$^{21}$ 
\beq P^{\gamma}-P^{\alpha}=
\frac{2\sigma^{\alpha\gamma}_{i}}{h_{i}}\;\;\;(i=1,\ldots\la -1),\;\;\;
P^{\gamma}-P^{\alpha}=
\frac{2(\sigma^{\gamma\delta}_{\la}-\sigma^{\alpha\delta})}
{h_{\la}}. \eeq
\par Equation (6) applied to the crystal is the equivalent of
Laplace's equation applied to liquid.
Thus, just as for a droplet, one can expect to find a high
pressure within a small crystal. It is this pressure that is
the cause of the increase in the chemical potential 
within the crystal.

\par Note that the line tension contributions to the free energy of crystal
formation were omitted in the model for homogeneous crystal nucleation in the 
surface-stimulated mode$^{21,22}$ because they were assumed to be negligible compared to the volume and
surface contributions. However, this assumption may no longer be valid for heterogeneous crystal nucleation because the nucleus is now much smaller (compared to the homogeneously formed one) and hence the contributions of three-phase contact lines can be more important.$^{31-33}$

Making use of equations (5) and (6), one can rewrite eq.(2) as
\beq W^{imm}=\nu[\mu^{\gamma}(P^{\alpha},T)-\mu^{\alpha}(P^{\alpha},T)]
+\sum_{i=1}^{\lambda -1}\sigma^{\alpha\gamma}_{i}A^{\alpha\gamma}_{i}
+\sigma_{\la}^{\gamma\delta}A_{\la}^{\gamma\delta}
-\sigma^{\alpha\delta}A_{\la}^{\gamma\delta}+\tau^{\alpha\gamma\delta}L^{\alpha\gamma\delta}.\eeq 
In this equation, the first term represents the excess Gibbs free
energy of the molecules in the crystal
compared to their Gibbs free energy in the liquid state. This
term is related to the enthalpy of fusion
$\De h$ by (see, e.g., ref. 23)
\beq \mu^{\gamma}(P^{\alpha},T)-\mu^{\alpha}(P^{\alpha},T)=-\int_{T_{0}}^{T}
\De h \frac{dT'}{T'},\eeq 
where $T_{0}$ is the melting temperature of the bulk solid ($T<T_0$), and $\De h<0$.

If the supercooling $T-T_0$ is not too large or, alternatively, if in the temperature
range between $T$ and $T_{0}$ the enthalpy of fusion does not change
significantly, eq.(8) takes the form
\beq \mu^{\gamma}(P^{\alpha},T)-\mu^{\alpha}(P^{\alpha},T)=-
\De h\ln\Theta.\eeq
with $\Theta=T/T_{0}$. 
Thus, one can rewrite eq.(7) in the following form
\beq W^{imm}=-\nu\De h\ln\Theta
+\sum_{i=1}^{\lambda-1}\sigma^{\alpha\gamma}_{i}A^{\alpha\gamma}_{i}+
\sigma_{\la}^{\gamma\delta}A^{\gamma\delta}_{\la}
-\sigma^{\alpha\delta}A_{\la}^{\gamma\delta}+\tau^{\alpha\gamma\delta}L^{\alpha\gamma\delta}.\eeq

\par By definition, the critical crystal (i.e., nucleus) is 
in equilibrium with the surrounding melt. For such a crystal
the first term in eq.(2) vanishes. 
On the other hand, for a crystal with one of
its facets being a crystal-foreign particle interface, and the others interfaced
with the liquid, one can show that
\beq V^{\gamma}(P^{\gamma}-P^{\alpha})=\frac2{3}\left(\sum_{i=1}^{\lambda-1}
\sigma^{\alpha\gamma}_{i}A^{\alpha\gamma}_{i}+
\sigma_{\la}^{\gamma\delta}A^{\gamma\delta}_{\la}
-\sigma^{\alpha\delta}A^{\gamma\delta}_{\la}\right).\eeq 
(This equality can be derived by representing $V^{\gamma}$ as the sum $\frac1{3}\sum_{i=1}^{\la}h_iA_i$ of the volumes of $\la$ pyramids with their bases at the crystal facets and their apexes at point O. The difference $P^{\gamma}-P^{\alpha}$ for every term in this sum is replaced by the RHS of the corresponding equality in eq.(6).) Substituting eq.(11) into eq.(2), one can thus obtain the following expression for the reversible work $W^{imm}_{*}$ of formation of a critical crystal: 
\beq W^{imm}_{*}=\frac1{3}\left(\sum_{i=1}^{\lambda-1}
\sigma^{\alpha\gamma}_{i}A^{\alpha\gamma}_{i}+
\sigma_{\la}^{\gamma\delta}A^{\gamma\delta}_{\la}
-\sigma^{\alpha\delta}A^{\gamma\delta}_{\la}\right)+
\tau^{\alpha\gamma\delta}L^{\alpha\gamma\delta},\eeq
or, alternatively, 
\beq W^{imm}_{*}=\frac1{2}V^{\gamma}_{*}(P^{\gamma}_{*}-P^{\alpha})+
\tau^{\alpha\gamma\delta}L^{\alpha\gamma\delta}\eeq
(hereafter the subscript ``*" indicates the quantities for the nucleus; it is omitted on the RHS of expressions for $W^{imm}_*$ to avoid the overcrowding of indices)
\par Clearly, in the atmosphere the crystal cluster forms not in the bulk
liquid, but within a liquid droplet (see Figure 1) which 
is itself surrounded by a vapor phase. The
reasoning here is almost identical to the preceding if, again, we neglect
the density difference between the liquid and crystal phases. One can easily show that all 
above equations, 
starting with eq.(2) and including eqs.(12) and (13) for
the reversible work $W^{imm}_{*}$ of formation of the critical crystal, remain
valid except that the pressures in the liquid and vapor phases are not equal but are 
related by the Laplace equation $P^{\alpha}=P^{\beta}+2\sigma^{\alpha\beta}/R$, with $R$ being
the radius of the droplet (assumed to remain constant during freezing).

\subsection{Foreign particle in contact with the liquid-vapor interface}

\par Now let us consider a foreign particle which is not immersed in a bulk liquid but is in some
kind of contact with the liquid-vapor interface (Figures 1 and 4), with the vapor phase being at
constant pressure and temperature. (Clearly, the particle must be completely wettable by the
liquid in order for the same particle to be able to be in mechanical and thermodynamic equilibrium
in both immersion and contact modes.) Now, upon sufficient supercooling, a crystal nucleus may
form heterogeneously with one of its facets (marked with the subscript ``$\lambda$'') on the
foreign particle and another one at the vapor-liquid interface. The latter facet will be marked
with the subscript $\la -1$.  All the other $\la-2$ facets lie within the liquid phase. 

\par 
Again, neglecting the density change upon freezing and the equality of pressures in the vapor and
liquid,  $P^{\alpha}=P^{\beta}$, the reversible  work of heterogeneous formation of the crystal
with its facet $\la$ on the foreign particle, the facet $\la-1$ interfacing vapor (i.e., in the
contact mode) and  all the others within the liquid phase  will be given by the expression 
\be W^{con}&=&\nu[\mu^{\gamma}(P^{\gamma},T)-\mu^{\alpha}(P^{\alpha},T)]-
V'^{\gamma}(P^{\gamma}-P^{\alpha})
+\sum_{i=1}^{\lambda-2}\sigma^{\alpha\gamma}_{i}A^{\alpha\gamma}_{i}
+\sigma_{\la-1}^{\beta\gamma}A_{\la-1}^{\beta\gamma} -\sigma^{\alpha\beta}A_{\la-1}^{\beta\gamma}
+\nonumber\\
&&\sigma_{\la}^{\gamma\delta}A_{\la}^{\gamma\delta} -\sigma^{\alpha\delta}A_{\la}^{\gamma\delta}
+\tau^{\alpha\beta\gamma}L^{\alpha\beta\gamma}
+(\tau^{\beta\gamma\delta}-\tau^{\alpha\beta\delta})L^{\beta\gamma\delta}
+\tau^{\alpha\gamma\delta}L^{'\alpha\gamma\delta}.\ee
Considering the contact mode, the prime will indicate quantities whereof the values may differ from
those in the immersion mode.  The equilibrium shape of the crystal (the Wulff form) is again
determined by  a series of equilibrium conditions on the  crystal ``edges'' formed by adjacent
facets.  For example,  the equilibrium conditions on the edges formed by the facets $\la-1$ and
$\la$ with the adjacent ones are given by
\beq \frac{\sigma^{\alpha\gamma}_{j}}{h_{j}}=
\frac{\sigma^{\beta\gamma}_{\la-1}-\sigma^{\alpha\beta}}{h'_{\la-1}}\;\;\;(j\ne\la),\;\;\;\;\;\;\;
\frac{\sigma^{\alpha\gamma}_{k}}{h_{k}}=
\frac{\sigma^{\gamma\delta}_{\la}-\sigma^{\alpha\delta}}{h_{\la}}\;\;\;(k\ne\la-1),\eeq
where $j$ and $k$ mark the facets adjacent to facets $\la-1$ and $\la$, respectively, and primes
will hereafter mark quantities for facet $\la-1$ at the droplet surface. Note again that the height
of the $\la-1$-th pyramid  (constructed with the base on facet $\la-1$ and with the apex at point
$O$ of the Wulff crystal) will differ from that with all of the facets (except for facet $\la$, see
Figure 3) in the liquid.  Thus, the shape of the crystal  will differ from that formed
heterogeneously in the immersion mode (i.e., when all facets, except for facet $\la$, are in
contact with the liquid. For this case, Wulff's relations take the form
\beq \frac{\sigma^{\alpha\gamma}_{1}}{h_{1}}=
\frac{\sigma^{\alpha\gamma}_{2}}{h_{2}}=\ldots
=\frac{\sigma^{\alpha\gamma}_{\lambda-1}-\sigma^{\alpha\beta}}{h'_{\lambda-1}
}=\frac{\sigma^{\alpha\gamma}_{\lambda}-\sigma^{\alpha\delta}}{h_{\lambda}
}.\eeq
Consequently, eq.(6) (the equivalent of Laplace's equation applied to crystals) becomes
\beq P^{\gamma}-P^{\alpha}=
\frac{2\sigma^{\alpha\gamma}_{i}}{h_{i}}\;\;\;(i=1,\ldots\la -2),\;\;\;
P^{\gamma}-P^{\alpha}=
\frac{2(\sigma^{\beta\gamma}_{\la-1}-\sigma^{\alpha\beta})}
{h_{\la-1}},\;\;\;
P^{\gamma}-P^{\alpha}=
\frac{2(\sigma^{\gamma\delta}_{\la}-\sigma^{\alpha\delta})}
{h_{\la}}. \eeq

Making use of equations (16) and
(17), one can rewrite eq.(14) as
\be W^{con}&=&\nu[\mu^{\gamma}(P^{\alpha},T)-\mu^{\alpha}(P^{\alpha},T)]
+\sum_{i=1}^{\lambda -2}\sigma^{\alpha\gamma}_{i}A^{\alpha\gamma}_{i}
+\sigma_{\la}^{\beta\gamma}A_{\la-1}^{\beta\gamma} -\sigma^{\alpha\beta}A_{\la-1}^{\beta\gamma}
+\sigma_{\la}^{\gamma\delta}A_{\la}^{\gamma\delta}-\sigma^{\alpha\delta}A_{\la}^{\gamma\delta}
+\nonumber\\
&&\tau^{\alpha\beta\gamma}L^{\alpha\beta\gamma}
+(\tau^{\beta\gamma\delta}-\tau^{\alpha\beta\delta})L^{\beta\gamma\delta}
+\tau^{\alpha\gamma\delta}L^{'\alpha\gamma\delta}
.\ee 
Furthermore, using eq.(9) one can represent eq.(18) in the following form
\be W^{con}&=&-\nu\De h\ln\Theta
+\sum_{i=1}^{\lambda-2}\sigma^{\alpha\gamma}_{i}A^{\alpha\gamma}_{i}
+\sigma_{\la-1}^{\beta\gamma}A^{\beta\gamma}_{\la-1}-\sigma^{\alpha\beta}A_{\la-1}^{\beta\gamma}
+\sigma_{\la}^{\gamma\delta}A^{\gamma\delta}_{\la} -\sigma^{\alpha\delta}A_{\la}^{\gamma\delta}
+\nonumber\\
&&\tau^{\alpha\beta\gamma}L^{\alpha\beta\gamma}
+(\tau^{\beta\gamma\delta}-\tau^{\alpha\beta\delta})L^{\beta\gamma\delta}
+\tau^{\alpha\gamma\delta}L^{'\alpha\gamma\delta}
.\ee
\par For a crystal with one of
its facets being a solid-vapor interface, and the others interfaced
with the liquid, one can show that
\beq V'^{\gamma}(P^{\gamma}-P^{\alpha})=\frac2{3}\left(\sum_{i=1}^{\lambda-2}
\sigma^{\alpha\gamma}_{i}A^{\alpha\gamma}_{i}
+\sigma_{\la-1}^{\beta\gamma}A^{\beta\gamma}_{\la} -\sigma^{\alpha\beta}A^{\beta\gamma}_{\la-1}
+\sigma_{\la}^{\gamma\delta}A^{\gamma\delta}_{\la} -\sigma^{\alpha\delta}A^{\gamma\delta}_{\la}
\right),\eeq
which makes it possible to represent the reversible work $W'_{*}$ of formation
of a critical crystal by the expression
\be W^{con}_{*}&=&\frac1{3}\left(\sum_{i=1}^{\lambda-2}\sigma^{\alpha\gamma}_{i}A^{\alpha\gamma}_{i}
+\sigma_{\la-1}^{\beta\gamma}A^{\beta\gamma}_{\la-1}-\sigma^{\alpha\beta}A^{\beta\gamma}_{\la-1}
+\sigma_{\la}^{\gamma\delta}A^{\gamma\delta}_{\la}-\sigma^{\alpha\delta}A^{\gamma\delta}_{\la}
\right)
+\nonumber\\
&&\tau^{\alpha\beta\gamma}L^{\alpha\beta\gamma}
+(\tau^{\beta\gamma\delta}-\tau^{\alpha\beta\delta})L^{\beta\gamma\delta}
+\tau^{\alpha\gamma\delta}L^{'\alpha\gamma\delta},\ee
or, alternatively, as
\beq W^{con}_{*}=\frac1{2}V'^{\gamma}_{*}(P^{\gamma}_{*}-P^{\alpha})
+\tau^{\alpha\beta\gamma}L^{\alpha\beta\gamma}
+(\tau^{\beta\gamma\delta}-\tau^{\alpha\beta\delta})L^{\beta\gamma\delta}
+\tau^{\alpha\gamma\delta}L^{'\alpha\gamma\delta}.\eeq
Equations (21) and (22) are similar to eqs.(12) and (13) which apply to heterogeneous
crystal nucleation in the immersion mode.
Along with eq.(13), equation (22) will be most useful in later discussions.

\par The reversible works of heterogeneous formation of crystal 
nuclei in the immersion and contact modes can now be compared. 
The difference between the internal pressure of the nucleus and the external pressure
does not depend on whether the nucleus forms in the immersed mode 
(let us denote it by $(P^{\gamma}_{*}-P^{\alpha})^{imm}$) or in the contact mode 
(denoted by $(P^{\gamma}_{*}-P^{\alpha})^{con}$). Indeed, by using
equation (9) and the equilibrium condition for the nucleus, namely
\beq \mu^{\gamma}(P^{\gamma},T)-\mu^{\alpha}(P^{\alpha},T)=0,\eeq
while assuming the crystal to be incompressible, one can show
that the difference $P^{\gamma}_{*}-P^{\alpha}$ for the
nucleus, in both cases, is determined by the supercooling
of the liquid, so that 
\beq (P^{\gamma}_{*}-P^{\alpha})^{con}=
(P^{\gamma}_{*}-P^{\alpha})^{imm}=\frac{\De h}{v}\ln\Theta ,\eeq
where $v$ is the volume per molecule in the crystal phase.
The first equality in eq.(24) is equivalent to
$$ \frac{\sigma^{\beta\gamma}_{\la-1}-\sigma^{\alpha\beta}}
{h'_{\la-1}}=\frac{\sigma^{\alpha\gamma}_{\la-1}}{h_{\la-1}}, $$
from which it follows that
\beq h'_{\la-1}=\frac{\sigma^{\beta\gamma}_{\la-1}-\sigma^{\alpha\beta}}
{\sigma^{\alpha\gamma}_{\la-1}}h_{\la-1}. \eeq
On the other hand,  $h_{i}=h'_{i}$ for $i=1,\ldots,\la -2,\la$,
by virtue of eqs.(6), (17), and  (24). This means that the Wulff
shape of the crystal, in the contact mode, is
obtained by simply changing the height of the $\la-1$-th pyramid of the Wulff
crystal in the immersion mode.
It is hence clear
that if $\sigma^{\beta\gamma}_{\la-1}-\sigma^{\alpha\beta} < \sigma^{\alpha\gamma}_{\la-1}$, 
then 
\beq h'_{\la-1}<h_{\la-1}\Rightarrow V'^{\gamma}_{*}<V^{\gamma}_{*}.\eeq 

According to eqs.(13) and (22), 
\beq
W^{con}_{*}-W^{imm}_{*}=\frac1{2}(V'^{\gamma}_{*}-V^{\gamma}_{*})(P^{\gamma}_{*}-P^{\alpha})+
\tau^{\alpha\gamma\delta}(L_{\la}^{'\alpha\gamma\delta}-L_{\la}^{\alpha\gamma\delta}) + 
(\tau^{\beta\gamma\delta}-\tau^{\alpha\beta\delta}) L_{}^{'\beta\gamma\delta}+
\tau^{\alpha\beta\gamma} L_{\la-1}^{\alpha\gamma\delta}. \eeq 
Because of eqs.(25) and (26), if 
\beq \sigma^{\beta\gamma}_{\la}-\sigma^{\alpha\beta} < \sigma^{\alpha\gamma}_{\la},\eeq
then the first term on the RHS of eq.(27) is negative. 

In the case of homogeneous crystal nucleation the line tension contributions to the free energy of
crystal nucleus formation  are either negligible or non-existent.$^{21,22}$  It allowed one to
conclude$^{21,22}$ that  if the condition in eq.(28)  is fulfilled,  it is thermodynamically more
favorable for the crystal nucleus to form with its facet $\la$ at the surface rather than within
the liquid.  Inequality (28) coincides with the condition of partial wettability of the $\la$-th
facet of the crystal by its own liquid phase.$^{23}$ This effect has been experimentally observed
for water-ice$^{26}$ at temperatures at or below 0$^{o}$C. In
those experiments,$^{26}$ when air was added to water vapor the partial wetting of ice by water
transformed into complete wetting, but {\bf only} for some orientations. Besides, the wettability
of solids by fluids usually decreases with decreasing temperature.$^{34,35}$ Since the freezing of
atmospheric water drops always occurs at temperatures far below 0$^{o}$C, one can expect the
partial wettability of at least some facets of water crystals even in the presence of air.
Furthermore, according to Cahn,$^{36}$ perfect wetting of a solid by a liquid away from the
critical point is not generally observed, i.e. the condition in eq.(28) should be fulfilled for
most substances. In Cahn's theory, the general restrictions on the solid phase are that its
surface is sharp on an atomic scale and interactions between surface and fluid are sufficiently
short-range. Therefore, that theory can be also applied to the case where the temperature is far
below the fluid critical point and the solid is of the same chemical nature as the fluid phases.
If the temperature approaches the fluid critical temperature, Cahn's theory becomes inapplicable. 
However, the temperatures involved in crystallization are usually far below the critical
point. All these combined with eq.(28) helps to explain why, in molecular dynamics simulation
studies, crystallization begins at or near a surface, and why it is easier, experimentally, to
observe the homogeneous crystallization of aerosols than that of the corresponding bulk liquid.

However, the presence of the line tension contribution on the RHS of eq.(27) makes it impossible to
draw unambiguous conclusions concerning the difference $W^{con}_{*}-W^{imm}_{*}$ for heterogeneous
crystal nucleation even when inequality (28) is fulfilled. Although the first term on the RHS of
eq.(27) is negative and gives rise to the thermodynamic propensity of the crystal nucleus to form
with the facet $\la-1$ at the droplet surface, the line tension contributions can be both negative
and positive because any of the  line tensions involved can be either negative or
positive.$^{31-33}$ Moreover, the sign of the line tension may change depending on
the temperature at which the crystallization occurs.

\section{Numerical evaluations and discussions}
To illustrate the above theory by numerical evaluations, consider first the homogeneous 
freezing of water droplets (surrounded by water vapor in air) at around $T_{hm}=233$ K 
(i.e., about $-40^o$C). The estimates for homogeneous crystal nucleation will serve as a
reference point to obtain some estimates of the relative importance of the line tension
contributions on the RHS of eq.(27) for $W^{con}_{*}-W^{imm}_{*}$.  

As reported by Defay {\em et al.}$^{23}$, the rate of homogeneous 
crystal nucleation in bulk supercooled 
water at this temperature is $7\times 10^{12} cm^{-3}s^{-1}$, with the nucleation
barrier height $W_*=45\;kT_{hm}$, the average (over all crystal facets) 
surface  tension of liquid-solid (water-ice) interface
$\sigma^{\alpha\gamma}$  being about $20$ dyn/cm (Table 18.1 in ref.23). The surface  tensions of
liquid-vapor and solid-vapor (ice-water vapor) interfaces at $T_{hm}=233$ K will be taken to be
$\sigma^{\alpha\beta}=88$ dyn/cm and $\sigma^{\beta\gamma}=103$ dyn/cm, respectively. 
All these values of $\sigma^{\alpha\gamma},\sigma^{\alpha\beta},$ 
and $\sigma^{\beta\gamma}$  are consistent with the data provided in ref.2. 


Let us assume that only the basal facets of the
hexagonal ice crystal is partially wettable by water at $T_{hm}=233$ K. The height of the
basal pyramid of the crystal cluster will be denoted by $\widetilde{h}_b$ 
when the basal facet is at the droplet 
surface (hereafter a tilde will mark quantities for this case)
and by $h_b$ when the entire crystal is immersed in the droplet.  
The corresponding works of homogeneous formation of crystal nuclei will be denoted as 
$\widetilde{W}_*$ and $W_*$. According to eqs.(31) and (32) of ref.21 and eqs.(15) and (16) of
ref.22, one can obtain for crystal nuclei:
$h_*\equiv 2h_{b}\simeq 21\times 10^{-8}\;\mbox{cm},\;\;
\widetilde{h}_*\equiv\widetilde{h}_{b}+h_{b}\simeq 18\times 10^{-8}\;\mbox{cm},\;\;
\widetilde{W}_*\simeq 39.3\,kT_{hm},\;\;
\De W_{*}\equiv \widetilde{W}_*-W_*\approx -5.7\; kT_{hm}$
 
These evaluations are for homogeneous crystal nucleation in the volume-based vs
surface-stimulated modes$^{21,22}$ which are equivalent to the immersion and contact modes,
respectively, of the heterogeneous crystal nucleation. At any particular temperature, the critical
crystal of heterogeneous nucleation is much smaller than for homogeneous nucleation. On the other
hand, the size of the nucleus increases with increasing temperature (i.e., decreasing supercooling).  Thus, one can expect that for any foreign particle there exists a temperature $233$ K $<T_{ht}<273$ 
K such that the linear size of the crystal nucleus (hence the number of molecules therein,
$\nu_c$) for heterogeneous nucleation is comparable to that estimated above for
homogeneous nucleation at $T_{hm}=233$ K. The energy unit $k_BT$ ($k_B$ is the Boltzmann constant) 
varies from $3.2\times 10^{-14}$ erg to $3.8\times 10^{-14}$ erg, i.e, by about $10$ \%. 

Let us assume, that for a selected foreign particle the temperature $T_{ht}\simeq 253$ K with the
thermal energy unit $k_BT_{ht}\simeq 3.5\times 10^{-14}$ erg. At this temperature, the first term on
the RHS of eq.(27) (hereafter referred to as the ``surface-stimulation term") can be roughly assumed
to be equal to $\De W_{*}$ because both quantities represent the surface contribution to the
difference between the free energy of nucleus formation in the surface-stimulated and volume-based
modes (for heterogeneous and homogeneous nucleation, respectively). Thus, according to the above
estimates,  
\beq \frac{\frac1{2}(V'^{\gamma}_{*}-V^{\gamma}_{*})(P^{\gamma}_{*}-P^{\alpha})}{k_BT_{ht}}
\approx\frac{\De W_{*}}{k_BT_{ht}}\approx 
-5.7\; \frac{T_{hm}}{T_{ht}}\approx - 5.3.\eeq 
As mentioned above, this contribution is negative if facet $\la-1$ (formed at the liquid-vapor
interface) is only partially wettable  by its melt (i.e., water), which is the case with the basal
facet of the crystals of hexagonal ice. Consequently, the droplet surface always makes the contact
mode of heterogeneous crystal nucleation in water droplets thermodynamically more favorable than
the immersion mode, regardless of the nature of the foreign particle. 

It is virtually impossible to provide general unambiguous estimates for the line tension
contributions to $W^{con}_{*}-W^{imm}_{*}$ in eq.(27). Indeed, the line tension is notorious not
only for the lack of reliable experimental data but (mostly) for its ability of being both negative
and positive and take values in the range from $10^{-1}$ to $10^{-5}$ erg (see, e.g.,
refs.31-33). Nevertheless, some estimates can provide useful insight into the problem
of ``contact mode vs immersion mode" of heterogeneous crystal nucleation. 

In the second term on the RHS of eq.(27), the difference
$L^{'\alpha\gamma\delta}-L^{\alpha\gamma\delta}$ represents the difference between the lengths of
the ``liquid-crystal-foreign particle" contact line in the contact and immersion modes. Clearly,
this difference is negative.  Considering, as above, that it is the basal facet of the hexagonal
ice crystal which forms at the liquid-vapor interface (with one of the six prismal facets formed on
the foreign particle), one can conclude that $L^{'\alpha\gamma\delta}-L^{\alpha\gamma\delta}\approx
-a = -8.65\times 10^{-8}$ cm for $q=2$ and
$L^{'\alpha\gamma\delta}-L^{\alpha\gamma\delta}\approx -a = -13.8\times 10^{-8}$ cm for $q=
0.5$. As for the line tension $\tau^{\alpha\gamma\delta}$, its sign can be expected to be
positive,$^{31-33}$ but we are not aware of any experimental or theoretical data
reported for  ``foreign particle-crystal-vapor" three-phase contact regions. Assuming that
$\tau^{\alpha\gamma\delta}$ can be anywhere in the range from $10^{-1}$ erg to $10^{-5}$ erg, it is
still most likely to be closer to $10^{-5}$ than to $10^{-1}$ erg because two out of three phases
in contact are solid phases involving little inhomogeneities of density profiles in the contact
region. Thus, one can cautiously suggest that: a) this three-phase contact line impedes the
heterogeneous crystal nucleation in the immersion mode vs contact mode; b) possible values of the
term $\tau^{\alpha\gamma\delta}(L^{'\alpha\gamma\delta}-L^{\alpha\gamma\delta})/k_BT$ may be
somewhere in the range from $-100$ to $-10$. 

The third term on the RHS of eq.(27) is due to the three-phase contact line ``liquid-vapor-foreign
particle". The length of this line, $L^{\beta\gamma\delta}$, is equal to
$-(L^{'\alpha\gamma\delta}-L^{\alpha\gamma\delta})$, evaluated in the above paragraph, i.e.,
$L^{\beta\gamma\delta}\approx8.65\times 10^{-8}$ cm for $q=2$ and $13.8\times 10^{-8}$ cm for
$q=0.5$. Further, the density inhomogeneities in the ``foreign particle-crystal-vapor" contact
region can be expected to be negligible compared to those in the ``foreign
particle-crystal-liquid", ``liquid-vapor-foreign particle", or ``liquid-vapor-liquid". Therefore,
one can consider that the line tension $\tau^{\beta\gamma\delta}$ is negligible when compared to
$\tau^{\alpha\beta\delta}$, so that the third term becomes 
$(\tau^{\beta\gamma\delta}-\tau^{\alpha\beta\delta})L^{\beta\gamma\delta}\simeq
-\tau^{\alpha\beta\delta}L^{\beta\gamma\delta}$. Depending on the wettability of the foreign
particle by liquid water in the water vapor, $\tau^{\alpha\beta\delta}$ can be positive as well as
negative. It was noted, however, that for the same foreign particle to be able to serve as an
equilibrium nucleating center in both immersion and contact modes it has to be completely wettable
by watter. Thus, one can suggest$^{27-29}$ that the line tension $\tau^{\alpha\beta\delta}<0$
with its absolute value closer to $10^{-5}$ than $10^{-1}$. Besides, the third
term on the RHS of eq.(27) can be expected to provide a contribution to $W^{con}_{*}-W^{imm}_{*}$ which 
is close to 
the contribution from the second term in absolute value and has an opposite sign. The approximate 
compensation of the second and third terms can thus be expected. 

The last term on the RHS of eq.(27) is due to the three-phase contact line ``liquid-vapor-crystal".
Considering again the basal facet of the hexagonal ice crystal forms at the liquid-vapor interface
(with one of the six prismal facets on the foreign particle), the length of this contact line is
approximately $L^{\alpha\beta\gamma}\approx 5 a$, that is, $L^{\alpha\beta\gamma}\approx 40.2\times
10^{-8}$ cm for $q=2$ and $L^{\alpha\beta\gamma}\approx 69.0\times 10^{-8}$ cm for
$q=0.5$. As the basal facet of an Ih crystal is partially wettable by liquid water with the
contact angle (measured inside the liquid phase) less than $\pi/2$, one can consider 
$\tau^{\alpha\beta\gamma}$ to be negative.$^{31,32}$ Even assuming for the value of
$\tau^{\alpha\beta\gamma}$ the smallest experimentally reported order of magnitude, $10^{-5}$ erg,
one can conclude that: a) this contact line significantly enhances the contact mode of
heterogeneous crystal nucleation compared to the immersion mode; b) the absolute value of the line
tension contribution to $W^{con}_{*}-W^{imm}_{*}$ from the ``vapor-liquid-ice" contact line is {\em
at least} by one order of magnitude greater than that of the surface-stimulation term (first term
on the RHS of eq.(27)). 
Thus, this line tension contribution to $W^{con}_{*}-W^{imm}_{*}$ can
dominate the surface-stimulation term. 

Evaluations for the case where the nucleus of an Ih crystal is formed a) 
in the immersion mode with the 
basal facet on the foreign particle and b) in the contact mode with the basal facet on the particle
and one of its prismal facet at the droplet-vapor interface can be carried out in a similar
fashion. Besides one can consider the case where in the contact mode one prismal facet forms on the
foreign particle and another at the droplet surface. Curiously, in this situation the foreign
particle does not even have to be in contact with the droplet surface. Moreover, the crystal
nucleus may form with one of its basal facets on the foreign particle and the other at the droplet
surface, and in this case, the foreign particle cannot be in contact with the droplet surface at
all (unless it has a very irregular, non-compact shape). In the latter case, the term `` contact
mode" is not even appropriate. Two common features of all these ``contact mode" situations are 
that: a) one of the crystal facets always forms at the droplet surface; b) there always
exists a contact ``vapor-liquid-crystal" of three water phases. Both of these factors (the latter
even significantly stronger than the former) thermodynamically favor the formation of a crystal
nucleus in the compact mode compared to the immersion one. The correctness of the term ``contact
mode" becomes, however, questionable, at least from a thermodynamic standpoint. One trivial
exception from the above consideration is the case where the surface of the foreign particle
touches the liquid-vapor interface from outside in parallel orientation (see Fig.1, case 4). In
this situation the same facet of the crystal nucleus forms at the droplet surface and on the
foreign particle, and there is no thermodynamic advantage for this mode compared to the immersion
mode (when the crystal nucleus forms with the same facet on the same surface of the foreign
particle).

It is worth emphasizing that for accurate calculations of $W_*^{imm}$ and $W_*^{con}$ it is necessary
to know not only the physico-chemical characteristics of the forming crystals (such as $\De h$, 
$\sigma$'s and $\tau$'s, etc...) but also the shape and size of the crystal nuclei. The latter,
however, can be accurately determined analytically if the former are known. 

Indeed, the shape of the crystal nucleus is determined by Wulff's relations (5) and (16). 
For example, since the shape of an ice crystal cluster is known (assumed to be a hexagonal prism), 
its state is completely determined by two geometric variables (provided that its density and
temperature are given), e.g., the height of the prism and the length of a side of a (regular) 
hexagon (the base of the prism). 
However, owing to Wulff's relations, eqs.(5) and (16), only one of these two variables is
independent. Therefore, both works $W^{imm}$ and $W^{con}$ are 
functions of only one independent variable, say, variable $a$, the length of a side of the hexagon. The
concrete form of the functions $W^{imm}=W^{imm}(a)$ and $W^{con}=W^{con}(a)$ depends on the mutual
orientation and location of the crystal cluster and foreign particle (and droplet surface in
case of $W^{con}$). 

For instance, consider a crystal cluster formed with one of its basal facets on a foreign particle
in the immersion mode. For the contact mode,  let us consider the same basal facet on the foreign
particle and a prismal facet (assumed to be only partially wettable by water) at the droplet
surface. Mark the basal facets with subscripts $1$ and $8$ and  the prismal facets with subscripts
$2,...,7$ (Figure 5). 

As agreed upon above, facet $8$ forms on the foreign particle. Clearly, in the immersion mode 
$\sigma^{\alpha\gamma}_p\equiv\sigma^{\alpha\gamma}_2=\cdots=\sigma^{\alpha\gamma}_7$,  
$A^{\alpha\gamma}_p\equiv A^{\alpha\gamma}_2=\cdots=A^{\alpha\gamma}_7$, 
$A^{\alpha\gamma}_b\equiv A^{\alpha\gamma}_1=A^{\alpha\gamma}_8$. In the contact mode the prismal
facet $7$ (assumed to be only partially wettable by liquid water) 
represents the crystal-vapor interface, hence 
$\sigma^{\alpha\gamma}_p\equiv\sigma^{\alpha\gamma}_2=\cdots=\sigma^{\alpha\gamma}_6$, 
$\sigma^{\beta\gamma}_p=\sigma^{\beta\gamma}_7$. 
Unlike the crystal cluster in the immersion mode, the basal facet in the contact mode is not a regular hexagon, 
$A^{'\alpha\gamma}_b\equiv A^{'\alpha\gamma}_1=A^{'\alpha\gamma}_8$ and 
$A^{'\alpha\gamma}_b< A^{\alpha\gamma}_b$, i.e., the surface areas of the basal facets in the
contact mode is smaller than that in immersion mode, according to eq.(25).  
Let us mark two prismal facets adjacent to facet $7$ by 
subscripts $2$ and $6$. Clearly, 
$A^{\alpha\gamma}_p=A^{\alpha\gamma}_3=A^{\alpha\gamma}_4=A^{\alpha\gamma}_5$, 
$A^{'\alpha\gamma}_p\equiv A^{\alpha\gamma}_2=A^{\alpha\gamma}_6<A^{\alpha\gamma}_p$, 
$A^{\beta\gamma}_p\equiv A^{\beta\gamma}_7 > A^{\alpha\gamma}_7$ (both inequalities are again due to eq.(25)). 

Let us use $a_i$ and $a'_i\;\;\;(i=2,...,7)$ to denote the length of the edge formed by the basal facet with prismal facet $i$ in the immersion and contact modes, respectively. In the immersion mode the base is a regular hexagon, i.e., $a\equiv a_2=...=a_7$. As clear from eq.(25), in the contact mode 
$a'_2=a'_6<a,\;\;a'_7>a$, whereas $a'_3=a'_4=a'_5=a$. 

In the first term on the RHS's of eqs.(10) and (19) the number of molecules in the crystal cluster can be represented as $\nu=\rho^{\gamma}V^{\gamma}$ or $\nu=\rho^{\gamma}V^{'\gamma}$, respectively, where 
$\rho^{\gamma}$ is the number density of molecules in phase $\gamma$ (ice). The volume of an Ih  crystal  (shaped as a hexagonal prism) is equal to the product {\em ``height of the prism"$\times$``surface area of the base"}. In both the immersion and contact modes the surface area of the base (regular hexagon in the former and irregular in the latter) is proportional to $a^2$, although coefficients of proportionality are different. In both cases, the height of the prism, $h$, is linearly related to $a$ according to Wulff's relations (5) and (16), respectively. Thus, in both eqs.(10) and (19) 
$\nu\propto \rho^{\gamma}a^3$. Likewise, one can show that all the surface tension and line tension terms on the RHS's of eqs.(10) and (19) are proportional to $a^2$ and $a$, respectively.  
Therefore, the reversible works of formation of a crystal cluster in these modes 
can be written (tedious but simple algebra is omitted) as  
\beq W^{imm}(a)= -I_3a^3+I_2a^2+I_1a, \;\;\;\;\;\;W^{con}(a)= -C_3a^3+C_2a^2+C_1a,\eeq
where $I_3,\;I_2,\;I_1$ and $C_3,\;C_2,\;C_1$ are positive coefficients,  
$$
I_3=\frac{9}{4}\rho^{\gamma}\De h \ln(\Theta) (\sigma^{\alpha\gamma}_b+\sigma^{\gamma\delta}_b-
\sigma^{\alpha\delta})/\sigma^{\alpha\gamma}_p,\;\;
I_2=\frac{3\sqrt{3}}{2}[2\sigma^{\alpha\gamma}_b+3(\sigma^{\gamma\delta}_b-\sigma^{\alpha\delta})],\;
I_1=6\tau^{\alpha\gamma\delta}, 
$$
and 
\be  
&&C_3=\rho^{\gamma}\De h \ln(\Theta) 
\left( \frac{3\sqrt{3}}{2}- C_h+\frac{C_h^2}{\sqrt{3}}\right) 
\frac{\sqrt{3}}{2}(\sigma^{\alpha\gamma}_b+\sigma^{\gamma\delta}_b-
\sigma^{\alpha\delta})/\sigma^{\alpha\gamma}_p,\nonumber\\ 
&&C_2=(\sigma_b^{\alpha\gamma}+\sigma_b^{\gamma\delta}-\sigma^{\alpha\delta})
\left(\frac{3\sqrt{3}}{2}- C_h+\frac{C_h^2}{\sqrt{3}}\right) +\nonumber\\
&&\left[(5-2(1-(\sigma_p^{\beta\gamma}-\sigma^{\alpha\beta})/\sigma_p^{\alpha\gamma}))
\sigma_p^{\alpha\gamma}+(\sigma_p^{\beta\gamma}-\sigma^{\alpha\beta})(2-
(\sigma_p^{\beta\gamma}-\sigma^{\alpha\beta})/\sigma_p^{\alpha\gamma})\right]
\frac{\sqrt{3}}{2}(\sigma^{\alpha\gamma}_b+\sigma^{\gamma\delta}_b-
\sigma^{\alpha\delta})/\sigma^{\alpha\gamma}_p,\nonumber\\
&&C_1=(\tau^{\beta\gamma\delta}-\tau^{\alpha\beta\delta}) 
(2-(\sigma_p^{\beta\gamma}-\sigma^{\alpha\beta})/\sigma_p^{\alpha\gamma})+
\tau^{\alpha\gamma\delta} 
(5-2(1-(\sigma_p^{\beta\gamma}-\sigma^{\alpha\beta})/\sigma_p^{\alpha\gamma}))+\nonumber\\
&&\tau^{\alpha\beta\delta} (2C_h+(2-(\sigma_p^{\beta\gamma}-\sigma^{\alpha\beta})/
\sigma_p^{\alpha\gamma})),
\nonumber
\ee
with 
$$
C_h=\frac{\sqrt{3}}{2}\left(1-\frac{\sigma^{\beta\gamma}_p-
\sigma^{\alpha\beta}}{\sigma_p^{\alpha\gamma}}\right)
$$ 
and $\rho^{\gamma}$ is the number density of molecules in the crystal phase. 

Using eq.(30), one can
find the length $a_*$ of a side of the hexagonal base of the crystal nucleus as the positive
solution of the equation $dW^{imm}(a)/da|_{a_*}=-3I_3a_*^2+2I_2a_*+I_1=0$, or alternatively, $dW^{con}(a)/da|_{a_*}=-3I_3a_*^2+2I_2a_*+I_1=0$.  which lead to 
$a_*=(2I_2+\sqrt{4I_2^2+12I_1I_3})/6I_3$ or $a_*=(2C_2+\sqrt{4C_2^2+12C_1C_3})/6C_3$ (see two paragraphs above eq.(30)). The height of the  crystal nucleus (shaped as a hexagonal 
prism)  is the same in both immersion and contact modes, 
$h_*=a_*\frac{\sqrt{3}}{2}(\sigma^{\alpha\gamma}_b+\sigma^{\gamma\delta}_b -
\sigma^{\alpha\delta})/\sigma^{\alpha\gamma}_p$. 

To numerically evaluate $a_*$ and $h_*$, information on  $\rho, \De h,
\sigma^{\alpha\beta}),\sigma^{\alpha\delta}),\sigma^{\alpha\gamma}_b,\sigma^{\alpha\gamma}_p, 
\sigma^{\gamma\delta}_b, \sigma^{\beta\gamma})_p $, $\tau^{\alpha\beta\gamma}, \tau^{\alpha\beta\delta}, 
\tau^{\alpha\gamma\delta}$, and $\tau^{\beta\gamma\delta}$ is needed.  
Experimental data on $\rho^{\gamma},\De h,$ and $\sigma^{\alpha\beta})$ are readily available (even as functions of temperature). For our evaluations they were taken to be 
$\rho^{\gamma}=0.92N_A/18$ cm$^{-3}$, 
$\De h\simeq 333.55\times 10^{7}N_A/18$ erg (where $N_A$ is the Avogadro constant), and $\sigma^{\alpha\beta})=83$ dyn/cm. Some data on $\sigma^{\alpha\gamma}_b$ and $\sigma^{\alpha\gamma}_p$ as well as on the mean value of the crystal-vapor surface tension have been also reported (see ref.2 for a short review). However, virtually no reliable 
data are currently available for any solid-ice interfacial tensions and line tensions in ``solid substrate-ice-liquid water-water vapor" systems. These were hence chosen somewhat arbitrarily, 
the main criterion being a reasonable agreement of the estimates extracted from the above equations with  those obtained from the experimental data on homogeneous ice nucleation. Considering nucleation of ice
crystals on a foreign particle such that at a given temperature  
$\sigma^{\alpha\delta}=40\; \mbox{dyn/cm},\;\; \sigma^{\alpha\gamma}_b=23\;
\mbox{dyn/cm},\;\; \sigma^{\alpha\gamma}_p=24\; \mbox{dyn/cm}, \sigma^{\gamma\delta}_b=50\;
\mbox{dyn/cm},\;\;\sigma^{\beta\gamma})_p =102\; \mbox{dyn/cm},\;\;\tau^{\alpha\beta\gamma}=-10^{-4}\; \mbox{dyn},\;\;\tau^{\alpha\beta\delta}=7\times 10^{-5}\; \mbox{dyn},\;\;\tau^{\alpha\gamma\delta}=10^{-5}\; \mbox{dyn}$, and 
$\tau^{\beta\gamma\delta}=5\times 10^{-6}\; \mbox{dyn}$ (reasonable choice according to scarce data available in literature), equations for $a_*$ and
$h_*$ would provide $a_*\simeq 29\times 10^{-8}$ cm, $h_*\simeq 35\times 10^{-8}$ cm, and $W_*^{con}-W_*^{imm}\simeq -12k_BT_{ht}$. 
As intended, these values are consistent with the estimates obtained above from the experimental data on
the homogeneous nucleation rate. 

\section{Concluding Remarks}

Previously, in the framework of CNT 
a criterion was found 
for when the surface of a droplet can  stimulate crystal nucleation therein so  that  the
formation of a crystal nucleus with one of its facets at the droplet surface is thermodynamically
favored (i.e., occurs in a surface stimulated mode) over its formation with all the facets {\em
within} the liquid phase (i.e., in a volume-based mode). For both unary$^{21}$ and
multicomponent$^{22}$ a
droplets,  this criterion  coincides with the  condition of partial wettability of at least one of
the crystal facets by  the melt (the contact angle, measured inside the liquid phase, must be greater than zero). 

However complex a theory of homogeneous crystal nucleation in droplets may be,  the presence of
foreign  particles,  serving as nucleating centers, makes the crystal nucleation phenomenon (and
hence its theory) even more involved. Numerous aspects of heterogeneous crystal nucleation still
remain obscure.  One of most intriguing problems in this field remains the strong enhancement of
heterogeneous crystallization in the contact mode compared to the immersion one. It 
has been observed that the same nucleating center initiates the
crystallization of a supercooled droplet at a higher temperature in the contact mode (with the
foreign particle just {\em in contact} with the droplet surface) compared to 
the immersion mode (particle {\em immersed} in the droplet).$^{2}$  
Many heterogeneous centers have different 
nucleation thresholds when they act in contact or immersion modes,  indicating that
the  mechanisms may be actually different for the different modes. 
Underlying physical reasons for this enhancement have remained largely 
unclear, but  the phenomenon of surface-stimulated (homogeneous)
crystal nucleation had strongly suggested  that  the droplet surface could enhance heterogeneous
nucleation in a  way similar to the enhancement of the homogeneous process. 

In this paper we have extended the approach, previously developed in refs.21,22, to heterogeneous
crystal nucleation on a  solid particle (in both immersion and contact modes) and have presented a
thermodynamic model shedding some light on the mechanism of the enhancement of this process in the
contact mode.  Our thermodynamic analysis suggests that the droplet surface can indeed
thermodynamically   enhance crystal nucleation in the contact mode compared to the immersion mode.
Whether this occurs  or not for a particular foreign particle is determined, however, by the
interplay between  various surface tensions and four line tensions involved in this process. As clear
from our model, the droplet surface may stimulate the heterogeneous crystal nucleation even in the
case where the foreign particle is actually completely immersed therein, but is situated closely
enough to the surface. This suggests that the term ``contact mode enhancement" is probably not very
appropriate for this phenomenon. 

As a numerical illustration of the proposed model, we have considered heterogeneous 
nucleation of Ih crystals on generic macroscopic foreign particles in water 
droplets at $T=253$ K. Our results suggest that while the droplet surface always
stimulates crystal nucleation on foreign particles in the ``contact mode", the 
line tension contribution to this phenomenon (due to the contact of 
three water phases, ``vapor-liquid-crystal") may be as important as the surface tension contribution.

\subsubsection*{}
{\em Acknowledgment} - {\small The authors thank F.M.Kuni and R.S.Kabisov for helpful discussions. 
}
\section*{References}
\begin{list}{}{\labelwidth 0cm \itemindent-\leftmargin}
\item $(1)$ {\it IPCC, Climate Change 2001: The scientific bases}. 
Inter government Panel on Climate Change; 
Cambridge University Press, Cambridge UK, 2001.
\item $(2)$ Pruppacher, H. R.; Klett., J. D. {\it Microphysics of clouds 
and precipitation}. (Kluwer Academic Publishers, Norwell, 1997).
\item $(3)$ Fletcher, N.H. {\it The physics of rainclouds}. (University Press, Cambridge, 1962).
\item $(4)$ Cox, S. K. {\it J. Atmos. Sci.} {\bf 1971}, {\it 28}, 1513.
\item $(5)$ Jensen, E. J.; Toon, O. B.; Tabazadeh, A.; Sachse, G. W.; Andersen,
B. E.; Chan, K. R.; Twohy, C. W.; Gandrud, B.; Aulenbach, S. M.;
Heymsfield, A.; Hallett, J.; Gary, B. {\it Geophys. Res. Lett.} {\bf 
1998}, {\it 25}, 1363.
\item $(6)$ Heymsfield, A.J.; Miloshevich, L.M. {\it J. Atmos. Sci.} 
{\bf 1993}, {\it 50}, 2335.
\item $(7)$ DeMott, P.J.; D. Cziczo, A. Prenni, D. Murphy, S. Kreidenweis,
D. Thomson, R. Borys, and D. Rogers, 2003a:  Proc. Nat. Acad. Sci.,
100, 14655-14660.
\item $(8)$DeMott, P.J.; Sassen, K., Poellot, M.; Baumgardner, D.; Rogers, D.;
Brooks, S.; Prenni, A.; Kreidenweis, S. {\it Geophys.
Res. Lett.} {\bf 2003}, {\it 30}, 1732, doi:10.1029/2003GL017410.
\item $(9)$Sassen, K.; P.DeMott, J. Prospero, and M. Poellot, 2003: 
Geophys. Res. Lett.,
30, 1633, doi:10.1029/2003GL017371.
\item $(10)$ Zuberi, B.; A. Bertram, T. Koop, L. Molina, and M.
Molina, 2001:  J. Phys. Chem. A, 105, 6458-6464.
\item $(11)$ Hung, H.-M.; A. Malinkowski, and S. Martin, 2003:
J. Phys. Chem. A, 107,
1296-1306.
\item $(12)$ Elias, V.; B. Simoneit, A. Pereira, J. Cabral, and J. Cardoso,
1999: 
Environ. Sci. Technol., 33, 2369-2376.
\item $(13)$ Popovitz-Biro, R.; J. Wang, J. Majewski, E. Shavit, L.
Leiserowitz, and M. Lahav, 1994: J. Amer. Chem. Soc., 116,
1179-1191.
\item $(14)$ Fukuta, N. Experimental studies of organic ice
nuclei. J.Atmos.Sci., {\bf 23}, 191-3196 (1966)
\item $(15)$ Vali, G. In Nucleation and Atmospheric Aerosols; Kulmala, M., 
Wagner, P., Eds.; Pergamon: New York, 1996. 
\item $(16)$ Shaw, R.A.; A.J.Durant, and Y.Mi, J.Phys.Chem.B {\bf 109}, 9865 (2005).
\item $(17)$ Fletcher, N.H., {\it J.Atmos.Sci.} {\bf 1970}, {\it 27}, 1098.
\item $(18)$ Guenadiev, N. {\it J.de Rech.Atmos.} {\bf 1970}, {\it 4}, 81.
\item $(19)$ Evans, L.F. {\it Preprints Conference Cloud Phys.}, Fort Collins, CO, p.14,
Am.Meteor.Soc., Boston.
\item $(20)$ Fukuta, N. {\it J.Atmos.Sci.} {\bf 1975}, {\bf 32}, 1597, 2371.
\item $(21)$ Djikaev, Y.S.; Tabazadeh, A.; Hamill, P.; Reiss, H. 
{\it J.Phys.Chem.} A {\bf 2002}, {\it 106}, 10247.
\item $(22)$  Djikaev, Y.S.; Tabazadeh, A.; Reiss, H. {\it J.Chem.Phys.} {\bf 2003}, {\it 118}, 
6572-6581.
\item $(23)$ Defay, R.; I. Prigogine, A. Bellemans, and D. H. Everett,
{\it Surface Tension and Adsorption} (John Wiley, New York, 1966).
\item $(24)$ Zell, J.; B. Mutaftshiev, Surf. Sci.
{\bf 12}, 317 (1968);
Grange, G.; Mutaftshiev, B. Surf. Sci. {\bf 47}, 723 (1975);
Grange, G.; R. Landers, and B. Mutaftshiev, Surf. Sci. {\bf 54}, 445 (1976).
\item $(25)$ Chatain, D.; and P.Wynblatt, in {\em Dynamics of Crystal Surfaces and Interfaces}, 
Ed. P.M.Duxbury and T.J.Pence, 53-58 (Springer, NY, 2002).
\item $(26)$ Elbaum, M.; S. G. Lipson, and J. G. Dash,
J. Cryst. Growth {\bf 129}, 491 (1993).
\item $^{27}$ Zasetsky, A.Y.; R. Remorov, and I.M. Svishchev, 
Chem.Phys.Let. {\bf 435} 50-53 (2007).  
\item $(28)$ Chushak, Y.G.; L.S.Bartell, J.Phys.Chem. B  {\bf 103}, 11196 (1999).
\item $(29)$ Rusanov, A.I. {\em Phasengleichgewichte und
Grenzflachenerscheinungen}; Akademie Verlag: Berlin, 1978.
\item $(30)$ Rowlinson, J.S.; Widom, B. {\it Molecular Theory of Capillarity} 
(Clarendon Press, Oxford, 1982).
\item $(31)$ Widom, B. J.Phys.Chem. {\bf 99}, 2803-2806 (1995).
\item $(32)$ Aveyard, R.; Clint, J.H. J.Chem.Soc., Faraday Trans. {\bf 92}, 85-89 (1996)
\item $(33)$ Amirfazli, A.; Neumann, A.W. Adv.Colloid Interface Sci. {\bf 110} 121-141 (2004).
\item $(34)$ Dietrich, S. In {\it Phase Transitions and Critical Phenomena}, Vol.12;
C. Domb and J. H. Lebowitz, Eds.; Academic Press: San Diego, 1988.
\item $(35)$ Sullivan, D. E.; Telo da Gama, M. M. In {\it Fluid
Intefacial Phenomena}; Croxton, C. A., Ed.; John Wiley \& Sons: New
York, 1986.
\item $(36)$ Cahn, J. W. {\it J. Chem. Phys.} {\bf 1977}, {\it 66}, 3667.
\item $(37)$ Libbrecht, K.G. Rep.Prog.Phys. {\bf 68}, 855-895 (2005).
\item $(38)$ Cahoon, A.; Maruyama, M.; Wettlaufer, J.S. Phys. Rev. Lett. {\bf 96} 255502 (2006).

\end{list}

\newpage
\subsection*{Captions} to  Figures 1 to 5 of the manuscript \\
{\sc ``Thermodynamics of heterogeneous crystal nucleation in 
contact and immersion modes"} by {\bf Y. S. Djikaev and E. Ruckenstein}

Figure 1. Heterogeneous crystal nucleation in a liquid droplet surrounded by vapor. 
``Immersion" mode: the crystal cluster forms  
with one of its facets on a foreign particle, completely immersed in a liquid droplet;ia  all
other crystal facets interface the liquid. ``Contact" mode: the foreign particle 
is in contact with the droplet surface; 
the cluster forms with one of the crystal 
facets on the particle, another facet 
at the liquid-vapor interface, and all other facets making the ``crystal-liquid" interface. Cases 1
through 4 represent a few of possible variations of the ``foreign particle--droplet surface" 
contact.
\vspace{0.1cm}\\
Figure 2. Illustration to Wulff's relations. 
The  surface area and surface tension of facet $i$ are denoted by $A_{i}$ and
$\sigma_{i}$, 
respectively; $h_{i}$ is the distance
from facet $i$ to reference point $O$.
\vspace{0.1cm}\\
Figure 3. Heterogeneous formation 
of a crystal nucleus on a foreign particle completely immersed in the liquid. 
\vspace{0.1cm}\\
Figure 4. Heterogeneous formation 
of a crystal nucleus on a foreign particle in contact with the liquid-vapor interface.  
\vspace{0.1cm}\\
Figure 5. Heterogeneous formation  of an Ih cluster on a foreign particle in the immersion and
contact modes.  contact with the liquid-vapor interface. The crystal cluster has a shape of a
hexagonal prism. One of the basal facets (facet 8) is formed on the foreign particle, the other
(facet 1) interfaces the liquid. Two of the prismal facets (with numbers 4 and 7) lie in the plane
perpendicular to the Figure. Prismal facets 5 and 6 cannot be seen by the reader, so they are shown
in the parentheses. 

\newpage
\begin{figure}[htp]
\begin{center}\vspace{1cm}
\includegraphics[width=8.3cm]{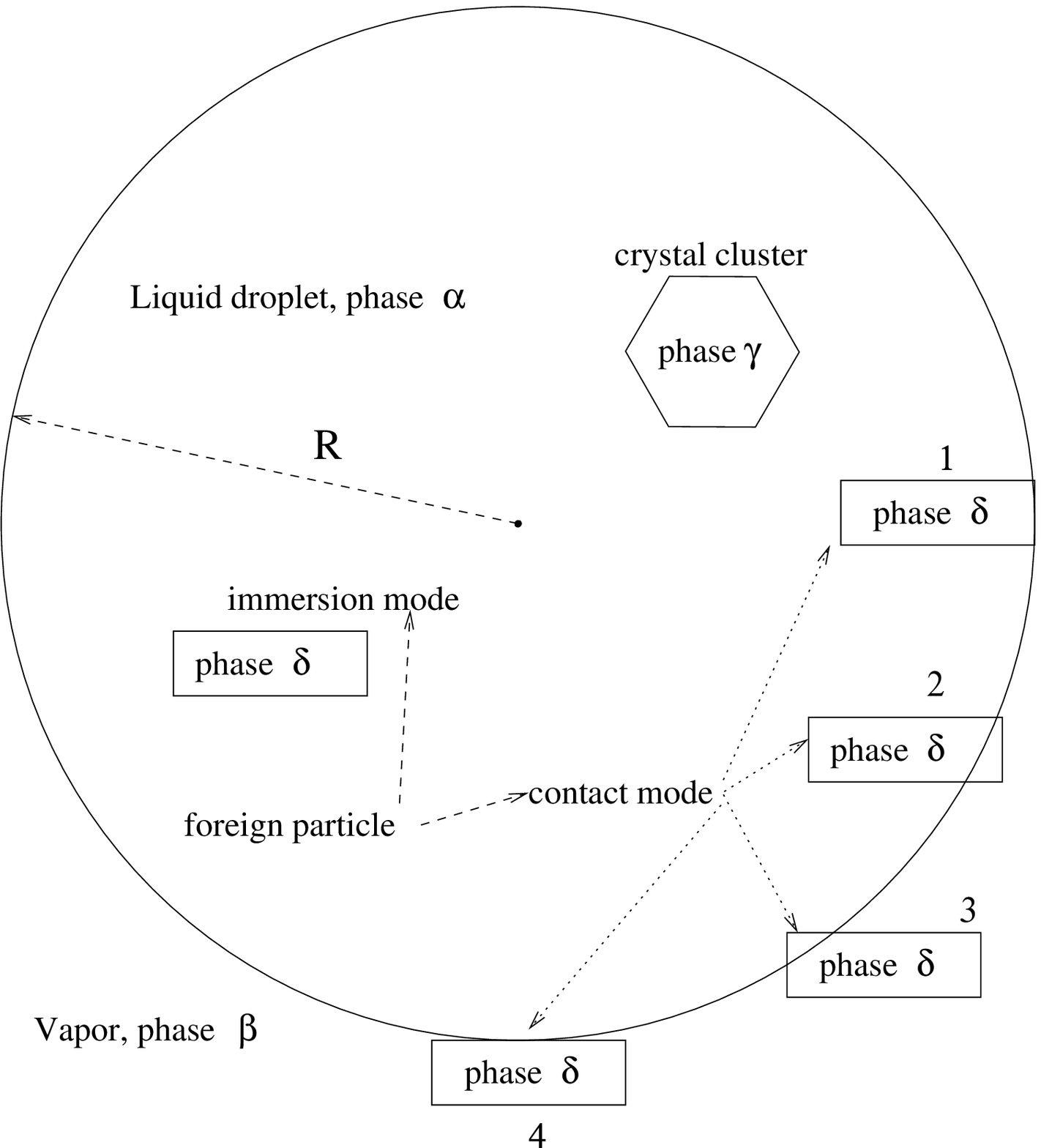}\\ [3.7cm]
\caption{\small }
\end{center}
\end{figure}

\newpage
\begin{figure}[htp]
\begin{center}\vspace{1cm}
\includegraphics[width=8.3cm]{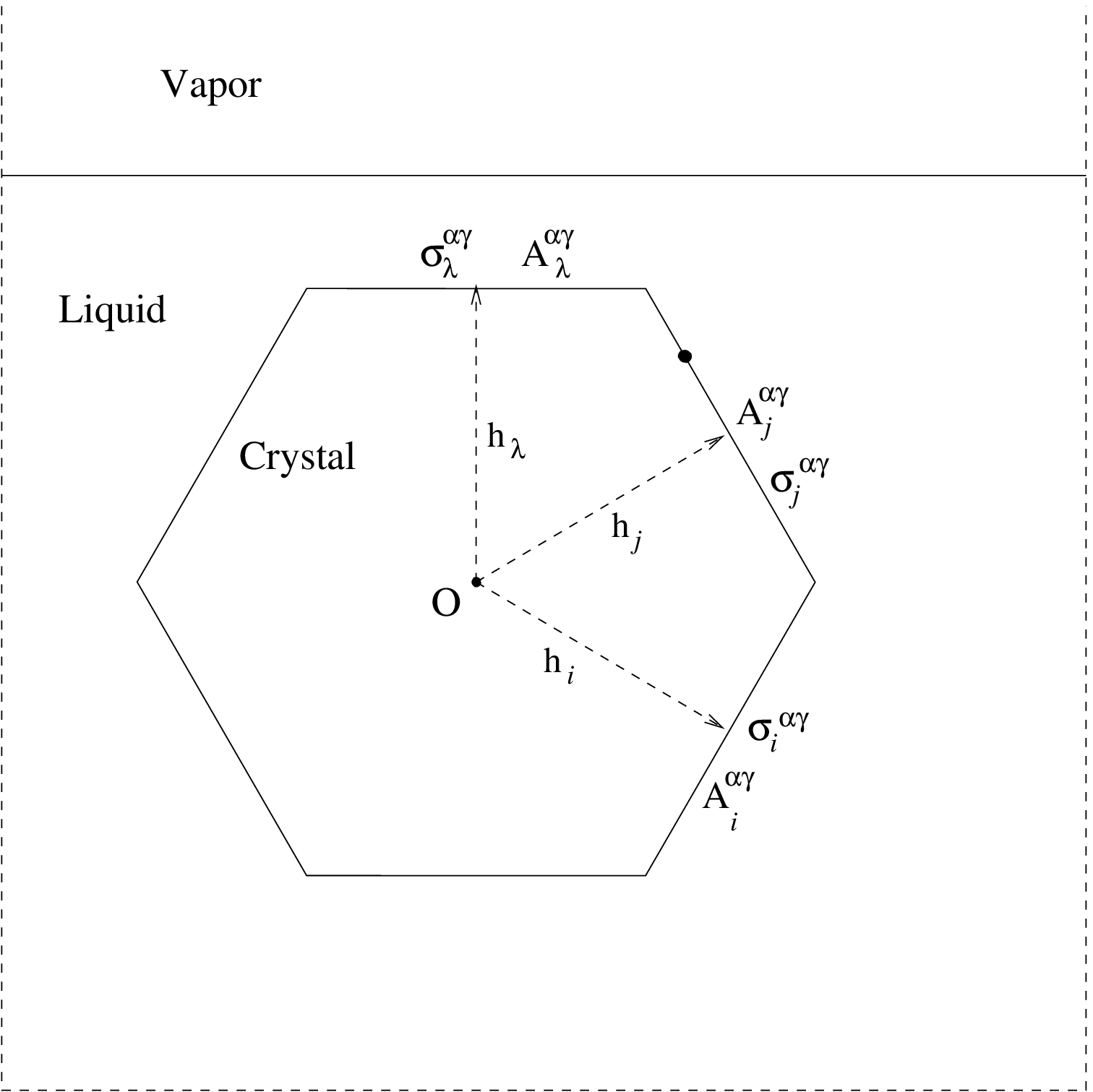}\\ [3.7cm]
\caption{\small }
\end{center}
\end{figure}

\newpage
\begin{figure}[htp]
\begin{center}\vspace{1cm}
\includegraphics[width=8.3cm]{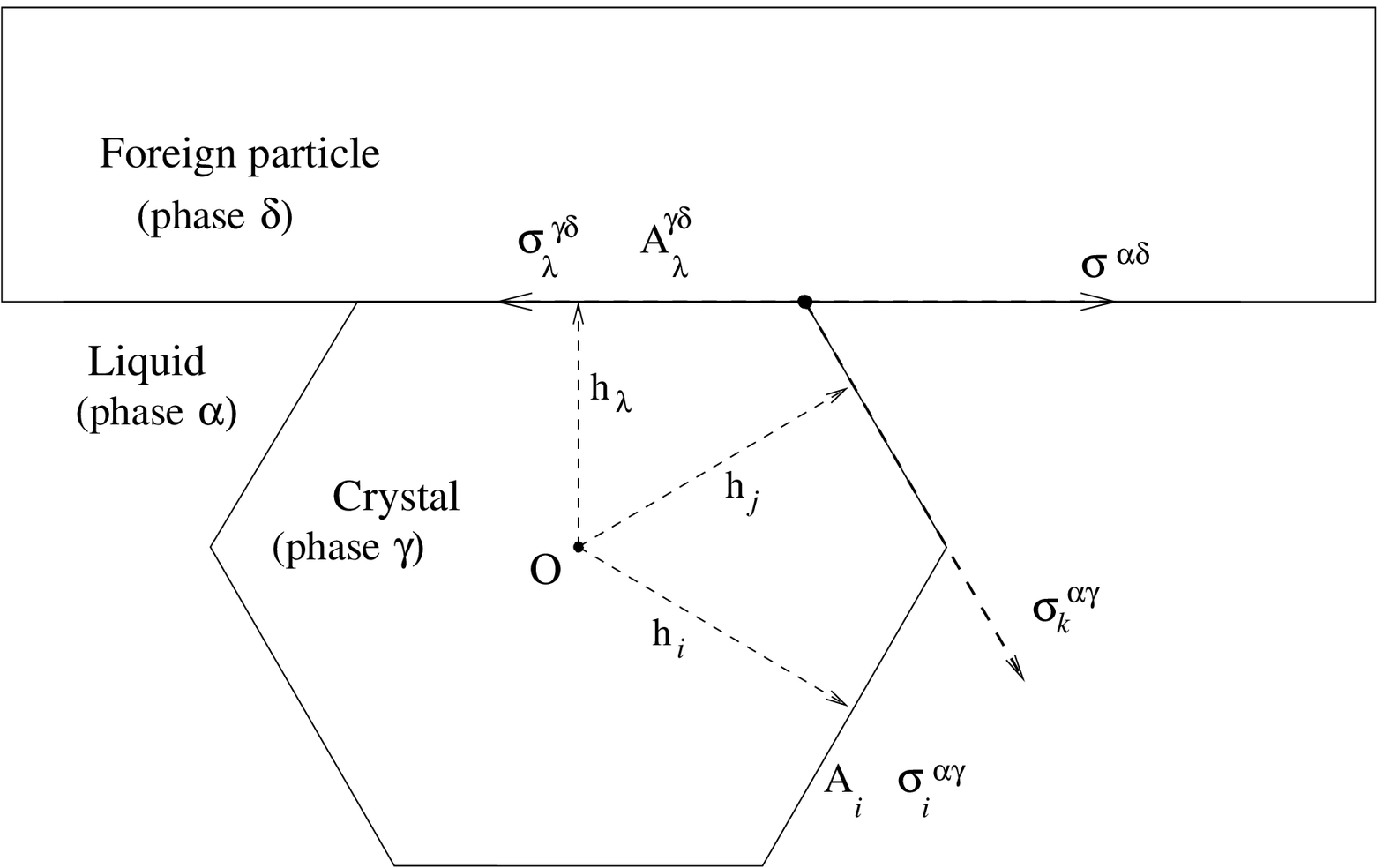}\\ [3.7cm]
\caption{\small }
\end{center}
\end{figure}

\newpage
\begin{figure}[htp]
\begin{center}\vspace{1cm}
\includegraphics[width=8.3cm]{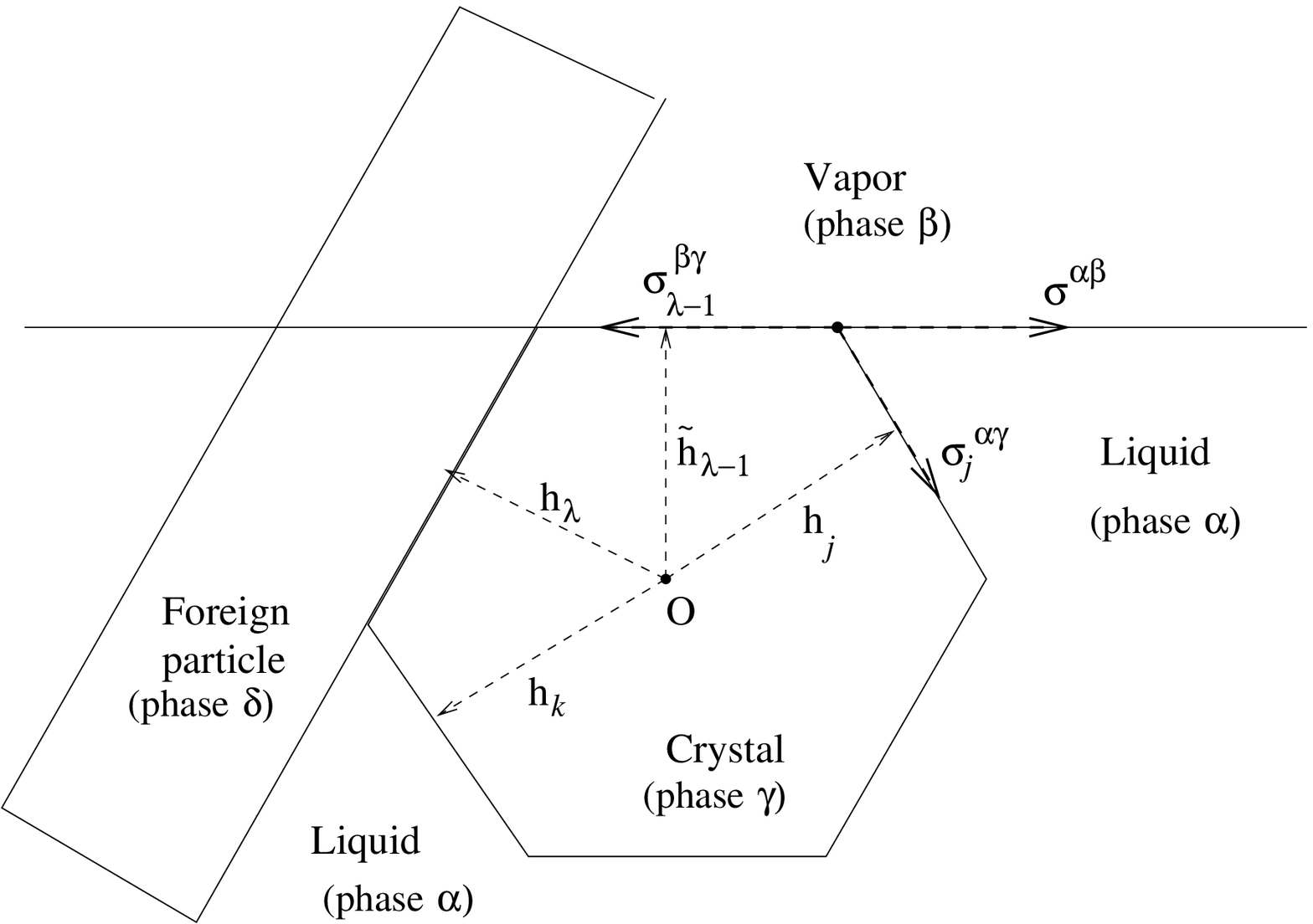}\\ [3.7cm]
\caption{\small }
\end{center}
\end{figure}

\newpage
\begin{figure}[htp]
\begin{center}\vspace{1cm}
\includegraphics[width=8.3cm]{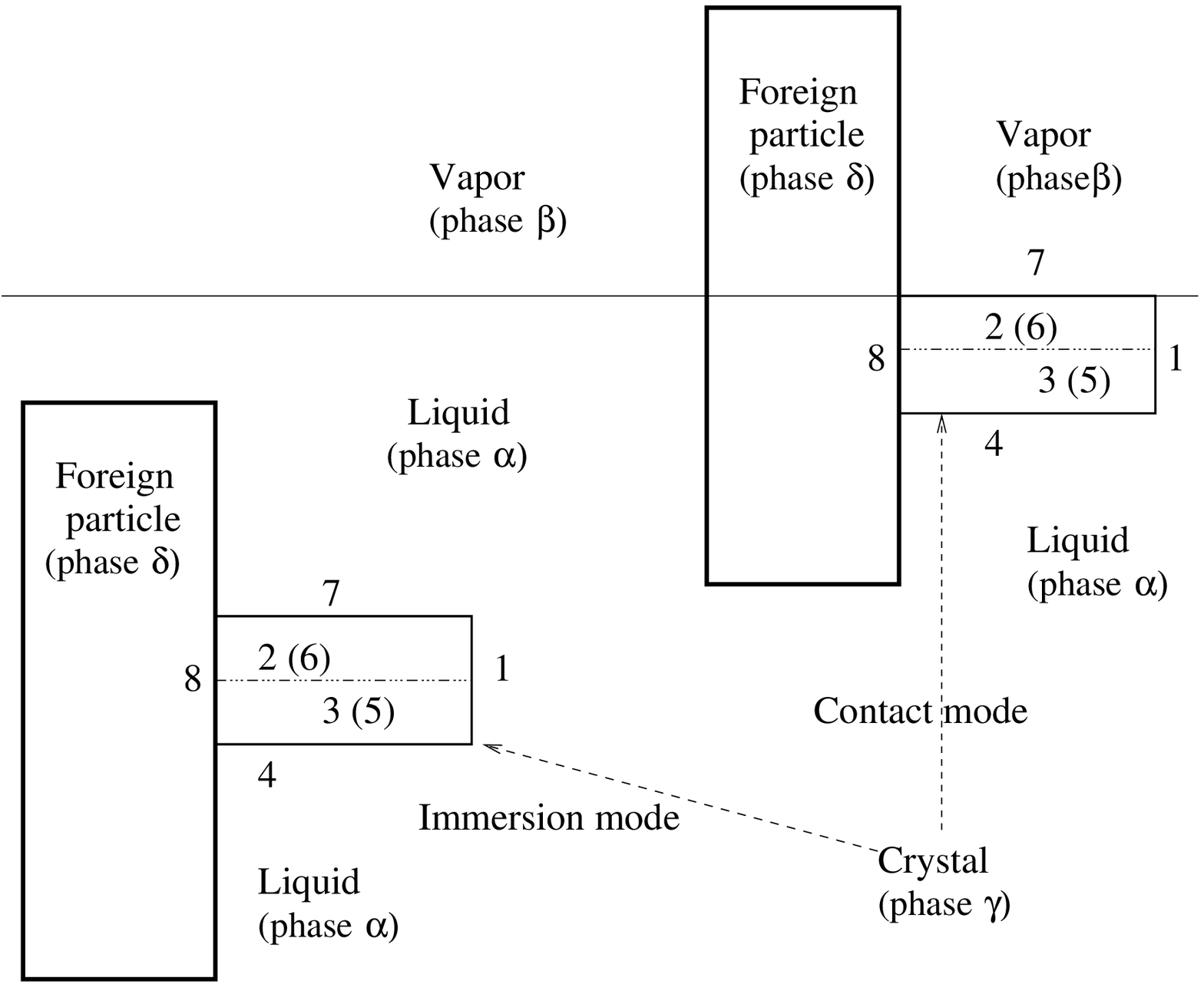}\\ [3.7cm]
\caption{\small }
\end{center}
\end{figure}

\end{document}